\documentclass[prd,showpacs,preprintnumbers,amsmath,amssymb]{revtex4}

\usepackage{graphicx}
\usepackage{dcolumn}
\usepackage{bm}

\begin{document}

\title{Schwinger Pair Production in $dS_2$ and $AdS_2$}

\author{Sang Pyo Kim}\email{sangkim@kunsan.ac.kr}
\affiliation{Department of Physics, Kunsan National University,
Kunsan 573-701, Korea}
\affiliation{Asia Pacific Center for
Theoretical Physics, Pohang 790-784, Korea}

\author{Don N. Page}\email{don@phys.ualberta.ca}
\affiliation{Theoretical Physics Institute, Department of Physics,
University of Alberta, Edmonton, Alberta, Canada T6G 2G7}

\date{2008 November 3}
\begin{abstract}
We study Schwinger pair production in scalar QED from a uniform
electric field in $dS_2$ with scalar curvature $R_{\rm dS} = 2 H^2$
and in $AdS_2$ with $R_{\rm AdS} = - 2 K^2$.  With suitable boundary
conditions, we find that the pair-production rate is the same
analytic function of the scalar curvature in both cases.
\end{abstract}

\maketitle

\section{Introduction}

Recently anti-de Sitter spacetime (AdS) has attracted much attention
because of the AdS/CFT correspondence.  The de Sitter spacetime (dS)
is also widely studied because of the accelerating phase of the
present Universe and the inflationary period in the early Universe.
The AdS and dS spacetimes have constant scalar curvatures and are
the maximally symmetric spacetimes for any given dimension.

The Minkowski vacuum becomes unstable with a strong electric field
and decays into pairs of charged particles, known as Schwinger pair
production \cite{Heisenberg-Euler,Weisskopf,Schwinger}.  The vacuum
also may be unstable when the spacetime expands or contracts,
leading to particle creation \cite{Parker,DeWitt}, particularly in
the de Sitter spacetime
\cite{Candelas,Lapedes,Mottola,Allen,BMS,Das-Dunne}. Schwinger pair
production in curved spacetimes may be an interesting issue,
combining the two effects. Pair production by a uniform electric
field in $dS_2$ was studied in Refs.
\cite{Garriga,Garriga2,Villalba} and in $AdS_2$ in Ref.
\cite{Pioline-Troost}.  Also pair production was considered in an
anisotropically expanding universe \cite{Villalba2,Villalba3}.
Though the instantons in $dS_2$ and $AdS_2$ have been known
independently, the relation between the pair-production rates has
not been examined before.

The purpose of this paper is to study scalar QED in a uniform
electric field in $dS_2$ and $AdS_2$ and to calculate the Schwinger
pair-production rates from the exact solutions of the massive
charged Klein-Gordon equation. Since the geometry of $dS_2$ with
scalar curvature $R_{\rm dS} = 2H^2$ can be analytically continued
to $AdS_2$ with $R_{\rm AdS} = - 2K^2$, both of which are
conformally flat, the Schwinger pair-production rates are shown to
have the same kind of analytical continuation from one geometry to
another.

The organization of this paper is as follows. In Sec. II, we
formulate the Klein-Gordon equation in a uniform electric field
in $dS_2$ and $AdS_2$. In Sec. III, we find the pair-production rate
from the exact mode solutions in $dS_2$. In Sec. IV, we find the
exact mode solutions and pair-production rate in $AdS_2$. In Sec. V,
we discuss the Schwinger pair-production rates and show that they
are both the same analytical function of the scalar curvatures in
both geometries.

\section{Klein-Gordon Equation in $dS_2$ and  $AdS_2$}

The two-dimensional de Sitter spacetime has positive constant
curvature and may be assigned the topology $R^1 \times S^1$, while
the anti-de Sitter spacetime has negative constant curvature and may
be assigned the topology $S^1 \times R^1$. The two-dimensional de
Sitter spacetime ($dS_2$) with a scalar curvature $R = 2H^2$ can be
embedded into a three-dimensional hyperboloid with radius $1/H$ and
has the metric \cite{Hawking-Ellis}
\begin{eqnarray}
ds_{\rm dS}^2 = - dt^2 + \cosh^2 (Ht) dx^2, \label{co-ds}
\end{eqnarray}
where $-\infty < t < \infty$ and $x$ is identified periodically with
period $2\pi/H$ to give the topology $R^1 \times S^1$, while the
two-dimensional anti-de Sitter spacetime ($AdS_2$) with scalar
curvature $R = - 2K^2$ has the metric of the form
\begin{eqnarray}
ds_{\rm AdS}^2 =  - \cosh^2 (Kx) dt^2 + dx^2, \label{co-ads}
\end{eqnarray}
where this time $- \infty < x < \infty$ and $t$ is identified
periodically with period $2 \pi /K$ to give the topology $S^1 \times
R^1$. Here $H$ and $K$ have the dimension of inverse length.
Alternatively, one may take $x$ to have infinite range in
(\ref{co-ds}) and $t$ to have infinite range in (\ref{co-ads}) to
get the covering spaces for $dS_2$ and $AdS_2$, respectively.

To study Schwinger pair production by a uniform electric field in
$dS_2$ and $AdS_2$, we consider scalar QED described by the
Klein-Gordon equation for bosons with mass $m$ and charge $q$ [in
units of $\hbar = c =1$]
\begin{eqnarray}
\Bigl[\frac{1}{\sqrt{-g}}(i \partial_{\mu} + qA_{\mu}) \Bigl(
\sqrt{-g} g^{\mu \nu} (i \partial_{\nu} + qA_{\nu}) \Bigr) + m^2
\Bigr] \Phi (t,x) = 0.
\end{eqnarray}
In a two-dimensional curved spacetime, a uniform electric field $E$
leads to the electromagnetic field two-form \cite{MTW}
\begin{eqnarray}
{\bf F} =  E \sqrt{|g|} dx \wedge dt.
\end{eqnarray}
Since the electromagnetic potential satisfies $d {\bf A} = {\bf F}$,
for $dS_2$ we may take the potential of the form
\begin{eqnarray}
{\bf A} =  - \frac{E}{H} \sinh (Ht) dx. \label{ds-pot}
\end{eqnarray}
Note that the potential (\ref{ds-pot}) respects one of the Killing
symmetries, $\partial_x$. In $AdS_2$ the electromagnetic potential
is given by
\begin{eqnarray}
{\bf A} =  \frac{E}{K} \sinh (Kx) dt. \label{ads-pot}
\end{eqnarray}
$AdS_2$ has the Killing vector, $\partial_t$, and allows separation
of variables.

\section{Pair Production in $dS_2$}

In the coordinates (\ref{co-ds}) for $dS_2$, the Klein-Gordon
equation minimally coupled with the potential (\ref{ds-pot}) takes
the form
\begin{eqnarray}
\Bigl[ \partial_{t}^2  + H \tanh (Ht) \partial_t + \frac{1}{\cosh^2
(Ht)} \Bigl(i \partial_x - \frac{qE}{H} \sinh (H t) \Bigr)^2 + m^2
\Bigr] \Phi (t, x) = 0.
\end{eqnarray}
Then the Fourier-component, $\Phi (t,x) = e^{ i k x} \phi_k
(t)/\sqrt{\cosh(Ht)}$, satisfies the one-dimensional equation
\begin{eqnarray}
[- \partial_{t}^2 + V_{\rm dS} (t) ] \phi_k (t) = 0, \label{ds-comp}
\end{eqnarray}
where
\begin{eqnarray}
V_{\rm dS} (t) =  - \frac{1}{\cosh^2 (Ht)} \Bigl(k + \frac{qE}{H}
\sinh (H t) \Bigr)^2 - m^2 + \frac{H^2}{4} \Bigl( 1 +
\frac{1}{\cosh^2 (Ht)} \Bigr).
\end{eqnarray}
In quantum mechanics, Eq. (\ref{ds-comp}) is a scattering problem
of a particle with a negative potential but with zero energy. In the
two asymptotic regions $t = \pm \infty$, there is an asymptotic
frequency
\begin{eqnarray}
\omega_0^2 = - V_{\rm dS} (\pm \infty) = \Bigl(\frac{qE}{H} \Bigr)^2
+ m^2 - \frac{H^2}{4},
\end{eqnarray}
so the positive frequency solutions of $\Phi_k (t)= \phi_k
(t)/\sqrt{\cosh (Ht)}$ at early and late times are given
asymptotically by
\begin{eqnarray}
u_{\rm in} (-\infty) \sim \frac{e^{- i \omega_0 t}}{\sqrt{2 \omega_0
\cosh (Ht)}},\quad u_{\rm out} (\infty) \sim \frac{e^{- i \omega_0
t}}{\sqrt{2 \omega_0 \cosh(Ht)}},
\end{eqnarray}
and the negative frequency solutions asymptotically by $u_{\rm
in}^*$ and $u_{\rm out}^*$. The initial vacuum and final vacuum are
defined with respect to $\Phi_{\rm in} (t, x) = e^{i kx} u_{\rm in}
(t)$ and $\Phi_{\rm out} (t, x) = e^{i kx} u_{\rm out} (t)$,
respectively. These are different definitions from the de Sitter invariant vacua \cite{Allen}, so our initial and final vacua are not the same, leading to pair production with respect to them.

From Ref. \cite{Po-Za} we find the general solution to Eq.
(\ref{ds-comp}) with two linearly independent solutions given by
\begin{eqnarray}
\phi_k (t) =   z^{n/2} (1-z)^{n^*/2} [c_1 F (\mu, \nu ; \gamma ; z)
+ c_2 z^{1 - \gamma} F (\mu - \gamma + 1 , \nu - \gamma + 1 ; 2-
\gamma ; z)],
\end{eqnarray}
where $F$ is the hypergeometric function \cite{Ab-St}, and $c_1$ and
$c_2$ are integration constants, and
\begin{eqnarray}
n&=& \frac{1}{2} - \frac{k}{H} + i \frac{qE}{H^2}, \nonumber\\
\mu &=&  \frac{n + n^*}{2}- i \frac{\omega_0}{H}, \quad
\nu = \mu^*, \nonumber\\
\gamma &=& n + \frac{1}{2}, \nonumber\\
 z &=& \frac{1 + i\sinh
(Ht)}{2}.
\end{eqnarray}
From the asymptotic formula \cite{formula} for $|z| \gg 1$, the
solution at early and late times can be written as
\begin{eqnarray}
\Phi_k (t) &=& D_1 u_{\rm in} (t) + D_2 u_{\rm in}^* (t), \nonumber\\
&=& D_3 u_{\rm out} (t) + D_4 u_{\rm out}^* (t),
\end{eqnarray}
where $D$'s are constants determined by $c$'s, $n$ and $\mu$ only.
Eliminating the $u^*_{\rm in}$-part and normalizing the remaining
part to $u_{\rm in} (t)$, for each momentum we obtain the frequency
mixing
\begin{eqnarray}
u_{\rm in} (t) = \alpha_k u_{\rm out} (t) + \beta_k u_{\rm out}^*
(t),
\end{eqnarray}
and the Bogoliubov transformation
\begin{eqnarray}
\hat{a}_{\rm out} (k) = \alpha_k \hat{a}_{\rm in} (k) + \beta^*_k
\hat{a}^{\dagger}_{\rm in} (k),
\end{eqnarray}
where \footnote{Phase factors are evaluated in the Riemann sheet
with $ 0 \leq {\rm arg} < 2 \pi$ so that $i = e^{i \pi/2}$ and $- i
= e^{i 3 \pi/2}$.}
\begin{eqnarray}
\beta_k = e^{i \pi (n^* - n  - 2\mu)/2}.
\end{eqnarray}
The coefficients satisfy the relation for bosons
\begin{eqnarray}
|\alpha_k|^2 - |\beta_k|^2 = 1.
\end{eqnarray}
Then the mean number of produced pairs with $k$ \cite{Parker,DeWitt}
\begin{eqnarray}
{\cal N}_k = \langle {\rm in} \vert \hat{a}^{\dagger}_{\rm out} (k)
\hat{a}_{\rm out} (k) \vert {\rm in} \rangle = |\beta_k|^2,
\end{eqnarray}
is given by the instanton-like `action', defined so that
$|\beta_k |^2 = e^{- {\cal S}_{\rm dS}}$, as
\begin{eqnarray}
{\cal S}_{\rm dS} &=& \frac{2 \pi}{H^2} \Bigl[ \sqrt{ (qE )^2 +
(mH)^2 - \frac{H^4}{4}} - qE \Bigr] \nonumber\\
&=& \frac{\pi m^2}{qE} \frac{2 - \frac{R}{4m^2}}{1 + \sqrt{1 +
\frac{m^2 R}{2 (qE)^2} - \frac{R^2}{16 (qE)^2}}}. \label{ex ins}
\end{eqnarray}
Here $R$ is the scalar curvature.

In the zero-electric field limit $(E = 0)$, we recover the
particle-production probability in de Sitter space
\cite{Candelas,Lapedes,Mottola,Allen,BMS,Das-Dunne}
\begin{eqnarray}
|\beta_k |^2 = e^{- \frac{2 \pi \sqrt{m^2 - \frac{H^2}{4}}}{H}},
\end{eqnarray}
which is the Boltzmann factor with the Gibbons-Hawking temperature
$T_{\rm GH} = H/ (2 \pi)$ when $H \ll m$. In the $H=0$ limit, we
recover the Schwinger pair-production rate in the two-dimensional
Minkowski spacetime. The instanton actions in Refs.
\cite{Garriga,Garriga2} are the limiting case of Eq. (\ref{ex ins})
when $H \ll m$. The pair-production rate of scalar QED may be
compared with that of spinor QED in Ref. \cite{Villalba}. A direct
calculation using the worldline instanton
\cite{Dunne-Schubert,Dunne-Wang,DWGS} and the WKB instanton action
\cite{Kim-Page3} also gives this limiting result,
with only the $-H^2/4$ term missing from the square root in the
first expression (\ref{ex ins}) above for ${\cal S}_{\rm dS}$.

\section{Pair Production in $AdS_2$}

In $AdS_2$ with the electromagnetic potential (\ref{ads-pot}), the
Fourier-component, $\Phi (t, x) = e^{- i \omega t} \varphi_{\omega}
(x)/\sqrt{\cosh (Kx)}$, satisfies the equation
\begin{eqnarray}
[ - \partial_x^2 + V_{\rm AdS} (x) ] \varphi_{\omega} (x) = 0,
\label{ads-comp}
\end{eqnarray}
where
\begin{eqnarray}
V_{\rm AdS} (x) = - \frac{\Bigl( \omega + \frac{qE}{K} \sinh (Kx)
\Bigr)^2}{ \cosh^2 (Kx)} + m^2 + \frac{K^2}{4} \Bigl( 1 +
\frac{1}{\cosh^2 (Kx)} \Bigr).
\end{eqnarray}
The formalism for $dS_2$ cannot be applied to this static problem
since there do not exist in-going states at past infinity
and out-going states at future infinity. However, we may
apply the tunneling idea for pair production: virtual pairs are
created from a tunneling barrier, which is the Dirac sea lowered by
the electric potential, and then move to spatial infinity to be real
pairs along the electric field
\cite{Keldysh,Nikishov-Ritus,Brezin-Itzykson,Narozhnyi-Nikishov,Popov,CNN,Kim-Page,Kim-Page2,Kim-Page3}.
For virtual pairs to be real ones, an asymptotic momentum given by
\begin{eqnarray}
k_0^2 = - V_{\rm AdS} (\pm \infty) = \Bigl(\frac{qE}{K} \Bigr)^2 -
m^2 - \frac{K^2}{4},
\end{eqnarray}
should be real, thus requiring $qE > \sqrt{(mK)^2 + K^4/4}$. Then
the solution takes asymptotically the form
\begin{eqnarray}
v_{\rm in}  (- \infty) \sim \frac{ e^{i k_0 x}}{\sqrt{2k_0 \cosh
(Kx)}}, \quad v_{\rm out}  (\infty) \sim \frac{ e^{i k_0
x}}{\sqrt{2k_0 \cosh (Kx)}}.
\end{eqnarray}
The $v_{\rm in}$ is the in-going wave to the barrier and $v_{\rm
out}$ is the out-going wave from the barrier.

As in the case of $dS_2$, we find the exact solution
\begin{eqnarray}
\varphi_{\omega} (t) =   z^{\tilde{n}/2} (1-z)^{\tilde{n}^*/2}
[\tilde{c}_1 F (\tilde{\mu}, \tilde{\nu} ; \tilde{\gamma} ; z) +
\tilde{c}_2 z^{1 - \tilde{\gamma}} F (\tilde{\mu} - \tilde{\gamma}
+1, \tilde{\nu} - \tilde{\gamma} +1; 2 - \tilde{\gamma}; z)],
\end{eqnarray}
where $\tilde{c}_1$ and $\tilde{c}_2$ are integration constants, and
\begin{eqnarray}
\tilde{n} &=& \frac{1}{2} - \frac{\omega}{K} + i \frac{qE}{K^2}, \nonumber\\
\tilde{\mu} &=&  \frac{\tilde{n}+ \tilde{n}^*}{2}- i \frac{k_0}{K},
\quad
\tilde{\nu} = \tilde{\mu}^*, \nonumber\\
\tilde{\gamma} &=& \tilde{n} + \frac{1}{2}, \nonumber\\
 z &=& \frac{1 + i\sinh
(Kx)}{2}.
\end{eqnarray}
Imposing the tunneling boundary condition by eliminating $v_{\rm
out}^*$ and appropriately normalizing the solution, we obtain
\begin{eqnarray}
v_{\rm out} (x) = \tilde{\alpha}_{\omega} v_{\rm in} (x) +
\tilde{\beta}_{\omega} v_{\rm in}^* (x),
\end{eqnarray}
where
\begin{eqnarray}
\tilde{\beta}_{\omega} = e^{i \pi (\tilde{n} - \tilde{n}^* + 2
\tilde{\mu})/2}.
\end{eqnarray}
The mean number can be expressed in terms of the instanton-like
`action' from the exact solution
\begin{eqnarray}
{\cal N}_{\omega} = |\tilde{\beta}_{\omega}|^2 = e^{ - {\cal S}_{\rm
AdS}},
\end{eqnarray}
where
\begin{eqnarray}
{\cal S}_{\rm AdS} &=& \frac{2 \pi}{K^2} \Bigl[ qE -  \sqrt{ (qE)^2
-(mH)^2 - \frac{K^4}{4}} \Bigr] \nonumber\\
&=& \frac{\pi m^2}{qE} \frac{2 - \frac{R}{m^2}}{1 + \sqrt{1 +
\frac{m^2 R}{2 (qE)^2} - \frac{R^2}{16 (qE)^2}}}. \label{ads-ex ins}
\end{eqnarray}
The instanton `action' (\ref{ads-ex ins}) agrees with that obtained
from the one-loop effective action in Ref. \cite{Pioline-Troost}.

\section{Conclusion}

We have studied Schwinger pair production by a uniform electric
field in $dS_2$ and $AdS_2$. We solved the Klein-Gordon equation in
these curved spacetimes and found the pair-production rate by
appropriately imposing boundary conditions. A number of interesting
points have been observed.

First, there is an analytical continuation of the Schwinger
pair-production rate between $dS_2$ with $R_{\rm dS} = 2 H^2$ and
$AdS_2$ with the scalar curvature $R_{\rm AdS} = - 2 K^2$. In fact,
the exact results are invariant under the correspondence $K = i H$
and $\omega = i k$. This is because the metric and the
electromagnetic field for $dS_2$ is analytically continued to
$AdS_2$ under the transformations $t \leftrightarrow i x$ and $x
\leftrightarrow i t$ together with $K \leftrightarrow i H$. The wave
function, say the right-moving free wave, is properly transformed
from one space into another. This means that the pair production is
given by the same analytic function of the scalar curvature in both
cases.

Second, the pair-production rate does not depend on the frequency or
momentum. A physical interpretation is that the frequency or
momentum depends on the Lorentz reference frame, whereas the
spacetimes and electromagnetic fields are maximally symmetric, so
that the result can only depend on the invariants $m^2$, $qE$, and
$R$ (and actually only on their two ratios, by dimensional
analysis).  It is interesting to notice in both cases the expected
numbers $e^{- {\cal S}}$ are given by the instanton-like `action' of
the form
\begin{eqnarray}
{\cal S}_{\rm dS} &=& \frac{2 \pi}{H^2} \Bigl[ \sqrt{ (qE )^2 +
(mH)^2 - \frac{H^4}{4}} - qE \Bigr], \nonumber\\
{\cal S}_{\rm AdS} &=& \frac{2 \pi}{K^2} \Bigl[ qE -  \sqrt{ (qE)^2
-(mH)^2 - \frac{K^4}{4}} \Bigr],
\end{eqnarray}
both of which can be written in terms of the scalar curvature $R$ as
\begin{eqnarray}
{\cal S} =  \frac{\pi m^2}{qE} \frac{2 - \frac{R}{4 m^2}}{1 +
\sqrt{1 + \frac{m^2 R}{2 (qE)^2} - \frac{R^2}{16 (qE)^2}}}.
\end{eqnarray}

Third, given the same magnitude (but not sign) of the scalar
curvature, $|R| = 2 K^2 = 2 H^2$, and the strength of electric
field, one has in general ${\cal S}_{\rm AdS} > {\cal S}_{\rm dS}$
and hence a larger pair-production rate in $dS_2$ than in $AdS_2$.
The gravitational confinement within $AdS_2$ suppresses the
Schwinger pair production, and there is a minimal strength $E_0 =
\sqrt{(mK)^2 + K^4/4}/q$ to be able to produce pairs. Without the
electric field, the pair-production rate is that of a scalar field
in the de Sitter spacetime, but vanishes in the anti-deSitter
spacetime due to the boundary condition at asymptotic regions, as
expected.

\acknowledgements

The work of S.P.K. was supported by the Korea Research Foundation
Grant funded by the Korean Government (MOEHRD) (KRF-2007-C00167) and
the work of D.N.P. by the Natural Sciences and Engineering Research
Council of Canada (NSERC). S.P.K. appreciates support by NSERC
through the University of Alberta, and D.N.P. appreciates support by
Korea Science Engineering Foundation (KOSEF) grant funded by the
Korea government (MOST) (No. F01-2007-000-10188-0) through Kunsan
National University while each of us visited the other institute.


\begin{references}
\bibitem{formula} For $|z| \gg 1$, the asymptotic form 15.3.7 of the
hypergeometric function \cite{Ab-St} is used, where $F(\mu, \nu;
\gamma; z) = a (-z )^{- \mu} + b (-z)^{- \nu}$ with $a$ and $b$
being determined by $\mu$, $\nu$ and $\gamma$ only.
\end{references}
\end{document}